\documentclass[twocolumn,showpacs,prd]{revtex4}
\usepackage{graphicx}
\usepackage{dcolumn}
\usepackage{bm}
\begin{document}

\title{Horizon Thermodynamics and Gravitational Tension}
\author{A. Widom and J. Swain}
\affiliation{Physics Department, Northeastern University, Boston MA USA}
\author{Y. N. Srivastava}
\affiliation{Physics Department, University of Perugia, Perugia IT}

\begin{abstract}
We consider the thermodynamics of a horizon surface from the viewpoint of 
the vacuum tension $\tau =(c^4/4G )$. Numerically, 
$\tau \approx 3.026\times 10^{43}$ Newton. In order of magnitude, this is the 
tension that has been proposed for microscopic string models of gravity. 
However, after decades of hard work on string theory models of gravity, 
there is no firm
scientific evidence that such models of gravity 
apply empirically. 
Our purpose is thereby to discuss the 
gravitational tension in terms of the conventional Einstein general theory 
of relativity that apparently does explain much and maybe all of 
presently known experimental gravity data. The central result is that matter 
on the horizon surface is bound by the entropy-area law by tension in the closely 
analogous sense that the Wilson action-area law also describes a surface confinement.
\end{abstract}


\maketitle

\section{Introduction \label{intro}}

Several decades of difficult mathematical work on string theories of gravity 
have yielded no definitive results concerning experimental gravitational systems. 
Yet, conventional general relativity\cite{Weinberg:1972,Landau:1999} contains 
the notion of a {\em vacuum tension} \begin{math} \tau =(c^4/4G ) \end{math} whose
large magnitude \begin{math} \tau \approx 3.026\times 10^{43} \end{math} Newton, 
determines the weak strength of the gravitational interaction. The vacuum tension 
is thought to determine the maximum 
force\cite{Unruh:1976,Jacobson:1995,Gibbons:2002,Schiller:2004} that can be exerted 
on any material body 
\begin{equation}
F\le \left[\tau = \left(\frac{c^4}{4G}\right)\right], 
\label{intro1}
\end{equation}
with equality taking place on bodies confined to a horizon surface.

In what follows we shall discuss the gravitational tension in terms of the 
Einstein gravitational field equations. The action-area result\cite{Wilson:1979} 
of Wilson which follows from the much lower tension of strong interaction string 
fragmentation models may (or may not) follow from QCD field theory\cite{Srivastava:2001}. Strong interaction 
string-like confinement has not been proved in detail from QCD field theory. 
But the Wilson gravitational action-area result is rigorously true for gravitational 
matter confinement on horizons. In Euclidean field theory the result is the entropy 
area theorem well known to be valid on black hole horizons.  

In Sec.\ref{tension} the notion of gravitational tension will be reviewed. 
The Killing vector associated with local conservation of material energy is 
discussed in Sec.\ref{mec}. The action and entropy per unit area associated with 
gravitational horizon surfaces are computed in Sec.\ref{ae}. For a black hole 
horizon, the surface tension of the horizon is computed and all energy and entropy 
is confined to the horizon surface as discussed in Sec.\ref{en}.  In the concluding 
Sec.\ref{conc}, the general notion of gravitational tension and horizon confinement 
is reviewed. 

\section{Gravitational Tension \label{tension}}
 
To understand the notion of gravitational tension, one may start from 
the gravitational field equations,    
\begin{equation}
R_{\mu \nu}-\frac{1}{2} g_{\mu \nu} R = 
\left(\frac{8\pi G}{c^4}\right)T_{\mu \nu } .
\label{tension1}
\end{equation}
The gravitational vacuum tension may be defined as 
\begin{equation}
\tau = \frac{c^4}{4G} 
\ \ \ \ \ \big[\tau \approx 3.026\times 10^{43}\ {\rm Newton}\big].
\label{tension2}
\end{equation}
The physical meaning of the gravitational tension resides 
in the notion of gravitational stress; It is  
\begin{equation}
T^G_{\mu \nu}=-\left[\frac{\tau }{2\pi }\right]
\left(R_{\mu \nu}-\frac{1}{2} g_{\mu \nu} R\right). 
\label{tension3}
\end{equation}
Curvature induces gravitational stress. The Einstein field equations 
simply state that the {\em total stress}, gravitational plus material, 
is null; 
\begin{equation}
T^G_{\mu \nu}+T_{\mu \nu}=0. 
\label{tension4}
\end{equation}
The very large value of the gravitational tension in 
Eq.(\ref{tension2}) means that a very small curvature gives rise 
to a very large gravitational stress in Eq.(\ref{tension3}). Since the 
total stress is null, i.e. 
\begin{math} T^{\rm tot}_{\mu \nu}=T^G_{\mu \nu}+T_{\mu \nu}=0 \end{math}, 
the total energy and total momenta are locally conserved in virtue of  
\begin{math} 0=0 \end{math}. On the other hand, the matter stress obeys 
\begin{equation}
D^\mu T_{\mu \nu}=0,  
\label{tension5}
\end{equation}
wherein \begin{math} D \end{math} is a covariant derivative so that 
Eq.(\ref{tension5}) is not a separate conservation law unless there is a 
Killing vector condition. Let us consider this in more detail.

\section{Matter Energy Conservation \label{mec}}
 
The matter stress tensor has precisely one time-like eigenvector, 
\begin{equation}
v^\mu v_\mu =-c^2 , 
\label{mec1}
\end{equation}
wherein 
\begin{equation}
T_{\mu \nu}v^\nu =-\varepsilon v_\mu , 
\label{mec2}
\end{equation}
\begin{math} \varepsilon \end{math} is the scalar energy per unit volume 
and \begin{math} v^\mu \end{math} is the local {\em material velocity four vector} 
of the matter flow. Taking the covariant derivative of Eq.(\ref{mec2}), 
\begin{eqnarray}
D^\mu \big(T_{\mu \nu}v^\nu \big)=\big(D^\mu T_{\mu \nu}\big) v^\nu 
+T_{\mu \nu} D^\mu v^\nu , 
\nonumber \\   
D^\mu \big(T_{\mu \nu}v^\nu \big)=T_{\mu \nu} D^\mu v^\nu , 
\nonumber \\ 
-D_\mu (\varepsilon v^\mu) = T_{\mu \nu} D^\mu v^\nu ,
\label{mec3}
\end{eqnarray}
in virtue of Eqs.(\ref{tension5}) and (\ref{mec2}). From Eqs.(\ref{mec3}) one 
may prove the following  
\par \noindent 
{\bf Theorem:} {\em If the material velocity is a Killing vector obeying} 
\begin{equation}
D^\mu v^\nu +D^\nu v^\mu = 0, 
\label{mec4}
\end{equation}
{\em then a local material conservation law of energy exists of the form}
\begin{equation}
D_\mu (\varepsilon v^\mu) = 0\ \ \ \ \Rightarrow 
\ \ \ \ \frac{1}{\sqrt{-g}}\ \partial_\mu 
\left[\sqrt{-g}\ (\varepsilon v^\mu )\right]=0. 
\label{mec5}
\end{equation}
Eq.(\ref{mec5}) follows from Eqs.(\ref{mec3}) and (\ref{mec4}) and the 
stress tensor symmetry \begin{math} T_{\mu \nu} = T_{\nu \mu} \end{math}.

\section{Action and Entropy \label{ae}}

Let us now consider the action and entropy on a horizon area 
element from the viewpoint of a quantum loop. We presume a Killing 
vector field over the horizon surface as in Eq.(\ref{mec4}). In space-time, 
a loop consists of a particle going forward in time and an anti-particle   
going backwards in time as well as space-like sections. The motion of a 
particle in a loop is surely mot a classically allowed path. But in space-time, 
the loop action \begin{math} \oint_{\partial\Sigma} {\cal W}=\int_\Sigma d{\cal W} \end{math} is 
{\em virtual} and occurs only in amplitudes as 
\begin{math} e^{i{\oint {\cal W}}/\hbar } \end{math}. For a 
purely spatial loop, Euclidean field theory dictates an entropy 
\begin{math} {\cal S} \end{math} such that 
\begin{equation}
\left[\frac{\cal W}{\hbar }\right] \ \ \to 
\ \ -\left[\frac{\cal S}{k_B}\right] . 
\label{ae1}
\end{equation}
Finally, the quantum gravitational length scale \begin{math} \Lambda \end{math} 
is determined by 
\begin{equation}
\Lambda ^2=\left(\frac{\hbar G}{c^3}\right) \ \ \ \ \ \ 
\big[\Lambda \approx 1.616\times 10^{-33}\ {\rm cm}\big]  
\label{ae2}
\end{equation}
that represents a natural area for discussing a horizon surface. 

\begin{figure}[tp]
\scalebox {0.5}{\includegraphics{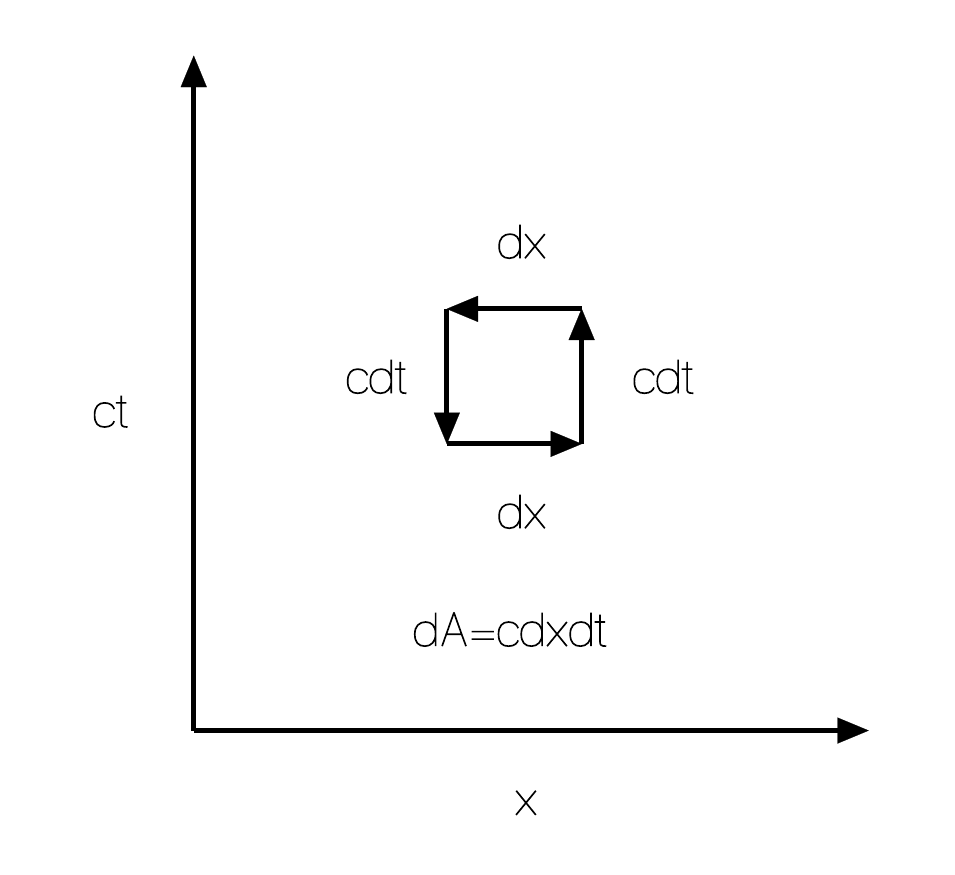}} 
\caption{A particle is shown schematically moving in a virtual closed 
infinitesimal loop in space-time. The area of the loop is given 
as a differential form by $dA=dx \wedge cdt$. }
\label{fig1}
\end{figure}

\subsection{Action \label{ac}} 

The action differential form for a displacement in the {\it x} direction is 
\begin{equation}
{\cal W}=p_x dx.  
\label{ac1}
\end{equation}
Employing the tension force \begin{math} (dp_x/dt)=\tau \end{math}, yields 
\begin{equation}
d{\cal W}=dp_x \wedge dx=\tau dt\wedge dx = 
- \left(\frac{\tau }{c} \right) dx \wedge cdt.  
\label{ac2}
\end{equation}
Thus, we find the Wilson area law for the gravitational tension on the horizon 
surface, i.e.  
\begin{equation}
\frac{d{\cal W}}{\hbar} =-\left(\frac{\tau }{\hbar c}\right)dA,
\label{ac3}
\end{equation}
as shown in FIG. \ref{fig1}. Note that 
\begin{equation}
\left(\frac{\tau }{\hbar c}\right)=\left(\frac{c^3 }{4 \hbar G}\right)
=\left(\frac{1}{4 \Lambda ^2}\right),
\label{ac4}
\end{equation}
yielding the Wilson area law
\begin{equation}
\frac{d{\cal W}}{\hbar} =-\left(\frac{dA}{4\Lambda ^2 }\right).
\label{ac5}
\end{equation}
The Euclidean field theory version of the area law follows from 
Eq.(\ref{ae1}).   

\subsection{Entropy \label{en}} 

From Eqs.(\ref{ae1}), (\ref{ac3}) and (\ref{ac5}) we go from the 
Wilson action area law for tension \begin{math} \tau \end{math} 
to the entropy area law on the horizon
\begin{equation}
\frac{d{\cal S}}{k_B} =\left(\frac{\tau }{\hbar c}\right)dA
=\left(\frac{dA}{4\Lambda ^2 }\right).
\label{en1}
\end{equation}
Equivalently, the entropy per unit area of horizon is given by 
\begin{equation}
\tilde{s}=\left(\frac{d{\cal S}}{dA}\right) =\left(\frac{k_B \tau }{\hbar c}\right)
=\left(\frac{k_B }{4\Lambda ^2 }\right).
\label{en2}
\end{equation}
Let us consider the meaning of the entropy-area law for a spherical black hole 
of radius 
\begin{equation}
R=\left(\frac{2GM}{c^2}\right)=\left(\frac{2G{\cal E}}{c^4}\right) 
=\left(\frac{\cal E}{2\tau}\right) ,
\label{en3}
\end{equation}
wherein \begin{math} {\cal E}=Mc^2  \end{math} is the black hole energy. The area 
of the horizon of the black hole is thereby 
\begin{equation}
A=4\pi R^2 =\pi \left(\frac{\cal E}{\tau }\right)^2 
\label{en4}
\end{equation}
with the entropy 
\begin{equation}
{\cal S}=\pi k_B \left(\frac{{\cal E}^2}{\hbar c \tau }\right), 
\label{en5}
\end{equation}
and temperature 
\begin{equation}
\frac{1}{T}=\frac{d{\cal S}}{d{\cal E}} \ \ \ \ \Rightarrow 
\ \ \ \ k_B T = 
\left(\frac{\tau }{2\pi }\right)\left(\frac{\hbar c }{\cal E }\right)=
\left(\frac{\hbar c}{4\pi R}\right).  
\label{en6}
\end{equation}
The free energy may be written as
\begin{equation}
{\cal F}={\cal E}-T{\cal S}=T{\cal S}=\frac{1}{2}{\cal E}.  
\label{en7}
\end{equation}
Finally, the surface tension of the black hole is 
\begin{math} \sigma ={\cal F}/A \end{math}; i.e. with 
\begin{math} \tilde{\varepsilon} ={\cal E}/A \end{math} 
\begin{equation}
\sigma =T\tilde{s}=\left(\frac{k_BT}{\hbar c}\right)\tau = 
\frac{1}{2}\tilde{\varepsilon}.    
\label{en8}
\end{equation}
Eq.(\ref{en8}) may also be written as the purely classical equation  
\begin{equation}
\sigma =\left[\frac{\tau }{4\pi R} \right]= 
\left[\frac{c^4 }{16\pi G R} \right] = 
\frac{1}{2}\tilde{\varepsilon}.    
\label{en9}
\end{equation} 
If one draws an equator around the sphere, then the resulting surface energy 
force \begin{math} 2\pi R\ \tilde{\varepsilon} = \tau \end{math} so that 
the two halves of the sphere attract each other via the vacuum 
gravitational tension \begin{math} \tau \end{math}. Eq.(\ref{en9}) is central 
to our discussion. The surface tension of the horizon is from the confinement 
of both energy and entropy on the surface of the black hole. There is no need 
to discuss what is ``internal'' to the black hole. All of the physical quantities 
are confined to the horizon surface by the gravitational tension.

\section{Conclusion \label{conc}} 

We have discussed the notion of gravitational tension and  the Killing vector 
associated with local conservation of material energy. The action and entropy 
per unit area associated with gravitational horizon surfaces have been computed 
as a Wilson loop with an action-area law employing gravitational tension. 
For a black hole  horizon, the surface tension of the horizon is computed. 
All energy and entropy is thereby confined to the horizon surface.  Horizon 
confinement is thereby proved to be a consequence of gravitational tension. There 
is {\em no need} to consider what happens on the {\em inside} of a black hole. 

\vfill

\section*{Acknowledgments}

J. S. would like to thank the United States National Science Foundation for support under PHY-1205845.


\begin{thebibliography}{14}

\bibitem{Weinberg:1972}
S. Weinberg, {\it Gravitation and Cosmology}, 
John Wiley \& Sons, New York (1972).  

\bibitem{Landau:1999}
L.D. Landau and E.M. Lifshitz, 
{\it The Classical Theory of Fields}, 
Butterworth-Heinemann, Amsterdam (1999). 

\bibitem{Unruh:1976}
W.G. Unruh, {\it Phys. Rev.} {\bf D14}, 870 (1976).

\bibitem{Jacobson:1995} 
T. Jacobson, {\it Phys. Rev. Lett.} {\bf 75}, 1260 (1995). 

\bibitem{Gibbons:2002}
G.W. Gibbons, {\it Foundations of Physics} 
{\bf 32}, 1891 (2002).

\bibitem{Schiller:2004}
C. Schiller, arXiv:physics/0309118v5 (2004).

\bibitem{Wilson:1979}
K.G. Wilson, {\it Phys. Rev.} {\bf 80}, 2445 (1979). 

\bibitem{Srivastava:2001}
Y. N. Srivastava, A. Widom 
{\it Phys.Rev.} {\bf D63} (2001) 077502\\
arXiv: hep-ph/0010064











\end{thebibliography}
\end{document}